\documentclass[
prl,preprint,
superscriptaddress,
showpacs,
amsmath,amssymb,
aps
]{revtex4-1}

\usepackage{graphicx}
\usepackage{amsmath}
\usepackage[version=3]{mhchem}
\usepackage[usenames]{color}
\usepackage{braket}
\usepackage{color}
\usepackage{braket}
\usepackage{textcomp}
\usepackage{physics}
\usepackage[colorlinks=true, linkcolor=blue, citecolor=blue, urlcolor=blue,pdftex]{hyperref}
\definecolor{newcolor}{rgb}{0.9,0,0.1}

\AtBeginDocument{\usepackage{booktabs}} 
\usepackage{multirow}
\usepackage{siunitx}

\usepackage{bm}

\newcommand{\figref}[1]{Fig.~\ref{#1}}
\newcommand{\unit}[2]{$#1\,\text{#2}$}
\newcommand{\equnit}[3]{$#1=#2\,\text{#3}$}

\begin{document}

\title{Vibronic response of a spin-1/2 state from a carbon impurity in two-dimensional WS$_2$}

\author{Katherine A. Cochrane}
\email[]{These authors contributed equally to this work.}
\affiliation{Molecular Foundry, Lawrence Berkeley National Laboratory, California 94720, USA}

\author{Jun-Ho Lee$^*$}
\affiliation{Molecular Foundry, Lawrence Berkeley National Laboratory, California 94720, USA}
\affiliation{Department of Physics, University of California at Berkeley, Berkeley, CA 94720, USA}
\author{Christoph Kastl}
\affiliation{Walter-Schottky-Institut and Physik-Department, Technical University of Munich, Garching 85748, Germany}

\author{Jonah B. Haber}
\affiliation{Molecular Foundry, Lawrence Berkeley National Laboratory, California 94720, USA}
\affiliation{Department of Physics, University of California at Berkeley, Berkeley, CA 94720, USA}

\author{Tianyi Zhang}
\affiliation{Department of Materials Science and Engineering, The Pennsylvania State University, University Park, PA 16082}
\affiliation{Center for Two-Dimensional and Layered Materials, The Pennsylvania State University, University Park, PA, 16802}

\author{Azimkhan Kozhakhmetov}
\affiliation{Department of Materials Science and Engineering, The Pennsylvania State University, University Park, PA 16082}

\author{Joshua A. Robinson}
\affiliation{Department of Materials Science and Engineering, The Pennsylvania State University, University Park, PA 16082}
\affiliation{Center for Two-Dimensional and Layered Materials, The Pennsylvania State University, University Park, PA, 16802}

\author{Mauricio Terrones}
\affiliation{Department of Materials Science and Engineering, The Pennsylvania State University, University Park, PA 16082}
\affiliation{Center for Two-Dimensional and Layered Materials, The Pennsylvania State University, University Park, PA, 16802}
\affiliation{Department of Physics and Department of Chemistry, The Pennsylvania State University, University Park, PA, 16802}

\author{Jascha Repp}
\affiliation{Institute of Experimental and Applied Physics, University of Regensburg, Regensburg, Germany}

\author{Jeffrey B. Neaton}
\email[]{jbneaton@lbl.gov }
\affiliation{Molecular Foundry, Lawrence Berkeley National Laboratory, California 94720, USA}
\affiliation{Department of Physics, University of California at Berkeley, Berkeley, CA 94720, USA}
\affiliation{Kavli Energy Nanosciences Institute at Berkeley, Berkeley, CA 94720, USA}

\author{Alexander Weber-Bargioni}
\email[]{afweber-bargioni@lbl.gov}
\affiliation{Molecular Foundry, Lawrence Berkeley National Laboratory, California 94720, USA}

\author{Bruno Schuler}
\email[]{bruno.schuler@empa.ch}
\affiliation{Molecular Foundry, Lawrence Berkeley National Laboratory, California 94720, USA}
\affiliation{nanotech@surfaces Laboratory, Empa -- Swiss Federal Laboratories for Materials Science and Technology, D\"ubendorf, Switzerland}

\begin{abstract}
We demonstrate the creation of a spin-1/2 state \textit{via} the atomically controlled generation of magnetic carbon radical ions (CRIs) in synthetic two-dimensional transition metal dichalcogenides (TMDs). Hydrogenated carbon impurities located at chalcogen sites introduced by chemical doping can be activated with atomic precision by hydrogen depassivation using a scanning probe tip. 
In its anionic state, the carbon impurity exhibits a magnetic moment of \unit{1}{$\mu_\text{B}$} resulting from an unpaired electron populating a spin-polarized in-gap orbital of C$^{\bullet -}_\text{S}$. Fermi level control by the underlying graphene substrate can charge and decharge the defect, thereby activating or quenching the defect magnetic moment. By inelastic tunneling spectroscopy and density functional theory calculations we show that the CRI defect states couple to a small number of vibrational modes, including a local, breathing-type mode. Interestingly, the electron-phonon coupling strength critically depends on the spin state and differs for monolayer and bilayer \ce{WS2}.
These carbon radical ions in TMDs comprise a new class of surface-bound, single-atom spin-qubits that can be selectively introduced, are spatially precise, feature a well-understood vibronic spectrum, and are charge state controlled.
\end{abstract}

\date{\today}
\pacs{}
\maketitle

The electron spin associated with atomic defects in crystals has been identified as a promising implementation of quantum bits (qubits)~\cite{weber2010quantum,awschalom2013quantum} that can be room temperature stable~\cite{maurer2012room}, optically addressable~\cite{gruber1997scanning}, and exhibit long coherence times~\cite{maurer2012room}. 
Nitrogen-vacancy centers in diamond~\cite{gruber1997scanning}, vacancies in SiC~\cite{koehl2011room}, donor atoms in silicon~\cite{schofield2003atomically}, and rare earth ions~\cite{xia2015all} are among the most prominent examples.
However, generating identical defects with the necessary atomic precision, designing them to be tunable by external fields, and often even knowing their exact identities have remained unsolved challenges in the field.\\

Two-dimensional (2D) materials, in particular semiconducting transition metal dichalcogenide (TMD) monolayers, promise to overcome these fundamental challenges~\cite{atature2018material,aharonovich2016solid}.
The large synthetic variability of TMDs enables bottom up impurity incorporation~\cite{das2015beyond} that facilitates engineering the defect electronic spectrum by chemical design principles. 
The monolayer confinement provides placement control in the out-of-plane direction and leads in general to deeper defect states, enhancing their atomic character.
The surface bound nature of prospective atomic qubits in TMDs immensely simplifies the coupling to external quantum circuits and offers new concepts for hetero-material integration~\cite{liu2019van}.
Most importantly, 2D TMDs enable charge state control~\cite{brotons2019coulomb}, spectral tuning by electrostatic gates~\cite{nguyen2019visualizing}, and high-fidelity electrical pumping schemes~\cite{Palacios-Berraquero2016,schuler2019electrically}. Furthermore, the dimensionality suppression of spin decoherence and the low abundance of non-zero nuclear spin isotopes in many TMDs is beneficial to increase the electron spin coherence time, superior in particular to boron nitride~\cite{de2003theory,ye2019spin}. Notably, the electron-phonon interaction, a defect-specific and potentially dominant source for decoherence in atomic-scale systems has yet to be adequately characterized~\cite{englund2010deterministic}.\\

Designing quantum emitters in 2D TMDs has been attempted by strain engineering~\cite{branny2017deterministic}, transmission electron microscopy~\cite{Lin2016DefectDichalcogenides}, and ion beam lithography~\cite{he2018defect,klein2019site}. However, the generated defects are often not identical, can only be placed with a limited spatial precision, and have mostly non spin-polarized states.
There has been extensive efforts to utilize boron nitride as a platform for quantum emitters, but the origin of reported behaviors in these materials is under debate~\cite{tran2016quantum,gottscholl2020initialization,Hayee2020revealing,mendelson2020identifying}.
Hence, a suitable, well-understood spin-polarized defect system in a 2D material that can be generated on-demand, with atomic precision and featuring favorable coherence properties has not yet been identified.\\

Here we introduce a carbon radical ion (CRI) in tungsten disulfide (\ce{WS2}) as a prototypical spin-1/2 two-level system that can be created with atomic precision while keeping the surrounding atomic structure virtually unchanged. Moreover, we demonstrate that the electron-phonon coupling associated with the two-level system is limited to a only few vibrational modes with coupling strengths similar to NV centers and exhibits a distinct spin and layer dependence.\\ 

Carbon impurity defects at chalcogen sites (C$_\text{X}$, X = S, Se) are created by scanning tunneling microscopy (STM)-induced hydrogen desorption from carbon-hydrogen (CH) complexes. Such CH impurities are frequently found in synthetic \ce{WS2} and \ce{WSe2}~\cite{cochrane2020intentional} but can also be deliberately created by post-synthetic methane plasma treatment, as shown here for \ce{WS2}~\cite{cochrane2020intentional,zhang2019carbon}.
We induce the hydrogen depassivation of CH$_\text{X}$ by a voltage pulse from the STM tip, which is highly reproducible and atomically precise. For \ce{WS2}, the Fermi level alignment with the graphene substrate results in a negatively charged carbon impurity with a radical character that we refer to as a CRI and denote as C$^{\bullet -}_\text{S}$. The CRI has an occupied spin-polarized defect state with a net magnetic moment of \unit{1}{$\mu_\text{B}$}. In \ce{WSe2}, we find C\textsubscript{Se} is neutral and thus has no magnetic moment. Moreover, we quantify the vibronic coupling of a single CRI to the host lattice by inelastic transport spectroscopy and using density functional theory (DFT) calculations. We find that the CRI two-level system couples predominantly to just a few vibrational modes. However, the coupling strength critically depends on the constituent state of the CRI two-level system as well as on the number of TMD layers.\\

\begin{figure*}[ht]
\includegraphics[width=\textwidth]{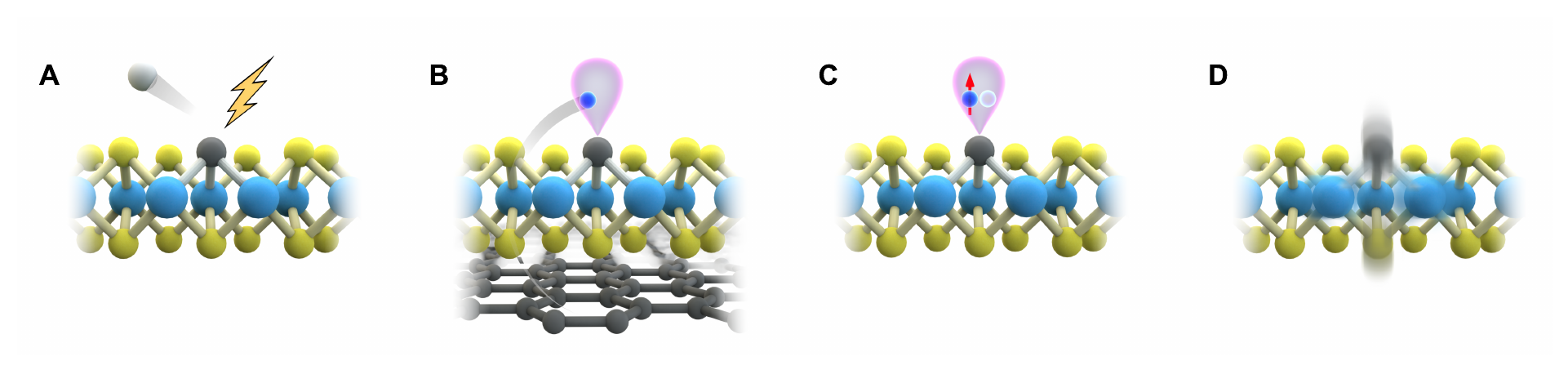}
\caption{\label{fig:TOC}
Schematic illustration of the hydrogen depassivation of the CH impurity (A), electron transfer from the substrate into the dangling C$_\text{S}$ bond (B), magnetic moment of the carbon radical ion (CRI) (C), and its vibronic coupling to the TMD host lattice (D).}
\end{figure*}

In the following we discuss our three primary conclusions: the hydrogen depassivation of the CH impurity and formation of the CRI, the two-level system associated with the CRI, and the vibronic coupling between the CRI two-level system with the TMD host lattice (see \figref{fig:TOC}).

\section*{Hydrogen Depassivation of a CH Impurity.}
\begin{figure*}[h]
\includegraphics[width=\textwidth]{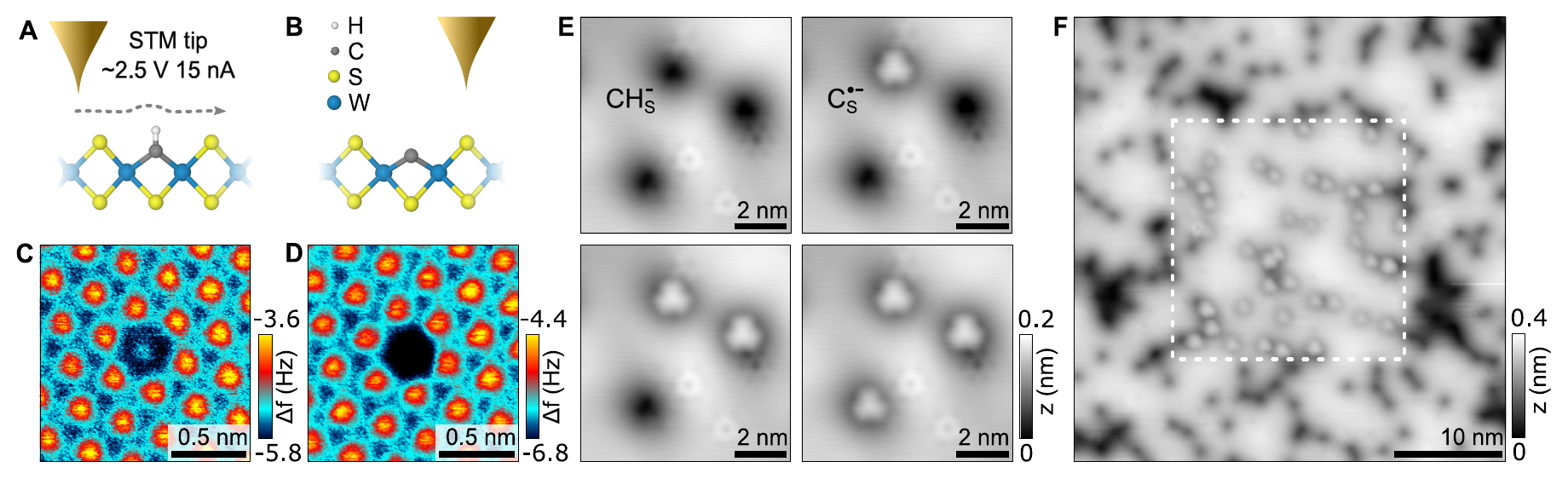}
\caption{\label{fig:Hdissociation}
\textbf{Hydrogen desorption from CH$_\text{S}$ impurities in deliberately CH-doped \ce{WS2}.}
\textbf{(A, B)} Atomic structure of a carbon-hydrogen complex (CH$_\text{S}$) and a carbon (C$_\text{S}$) substituent at a sulfur site in \ce{WS2}. (A) shows the pathway of the tip for the dehydrogenation process. \textbf{(C, D)} Constant height CO-tip nc-AFM images of the CH$_\text{S}$ (C) and C$_\text{S}$ (D) defect. \textbf{(E)} Sequential, controlled conversion of CH$^-_\text{S}$ to C$^{\bullet -}_\text{S}$ by H desorption (\equnit{V}{1.1}{V}, \equnit{I}{100}{pA}). The sample bias and current was increased to \unit{2.5}{V} and \unit{15}{nA} and rastered over an individual defect with images taken between each conversion. \textbf{(F)} Large scale image showing conversion of a center subsection (20$\times$20\,nm$^2$) outlined by the white dashed box (\equnit{V}{1.2}{V}, \equnit{I}{100}{pA}).
}
\end{figure*}

In \figref{fig:Hdissociation}, STM and CO-tip noncontact atomic force microscopy (nc-AFM) images of CH$^-_\text{S}$ defects in deliberately CH-doped monolayer \ce{WS2} (0.6\,\% atomic doping concentration) are presented. In the nc-AFM images, CH$^-_\text{S}$ defects appear as a small protrusion at a sulfur site (\figref{fig:Hdissociation}C), in excellent agreement with the simulated AFM contrast obtained from the relaxed geometry using DFT calculations~\cite{cochrane2020intentional}. In STM, CH$_\text{S}^-$ is imaged as a large, circular depression at positive bias resulting from upwards band bending due to the negative charge~\cite{aghajanian2020resonant,schuler2019substitutional}. After scanning the tip over the defect at high applied sample biases and high tunneling current set-points ($\sim$2.5\,V and $\sim$15\,nA), a dramatic change in the STM and AFM contrast is observed. In AFM, the small protrusion disappears (see \figref{fig:Hdissociation}D). In STM, a three-fold symmetric, bright orbital structure on a dark background is observed at positive voltage. 
Based on the prior knowledge of the precursor defect by targeted doping, the nc-AFM simulations, and the characteristic electronic fingerprint of the converted defect (discussed below), we show that the conversion process is a controlled desorption of the hydrogen atom from the CH complex.
The hydrogen desorption by the STM tip is likely a resonant process where tunneling into an unoccupied CH$_\text{S}$ defect state weakens the C-H bond. This defect state as identified by DFT calculations exhibits a local anti-bonding character with a nodal plane between the carbon and hydrogen atom, supporting this hypothesis~\cite{cochrane2020intentional}. Tunneling at negative bias with a comparable magnitude does not result in H-desorption.\\

Hydrogen desorption by STM has been reported for hydrogen-terminated silicon surfaces~\cite{shen1995atomic,schofield2003atomically,achal2018lithography} and organic molecules~\cite{schuler2013adsorption,van2013suppression}. As with these systems, removal of the hydrogen creates a dangling bond with a radical character, hence the defect becomes ``depassivated''. The dehydrogenation of the CH defect is very reliable and can be performed with single-atom precision as seen in \figref{fig:Hdissociation}E. Larger patterns can be written by scanning the surface at elevated bias and currents, shown in  \figref{fig:Hdissociation}F. 
Occasionally, a hydrogen atom reattaches from a previously depassivated C$_\text{S}$ defect while scanning at high bias (see Supplementary Materials, SM). We speculate that a H atom residing close to the tip apex transfers back and passivates the defect again, as previously suggested~\cite{pavlivcek2017tip}. This shows that the process is reversible.
Even more rarely, the entire CH complex is removed irreversibly, creating a sulfur top vacancy.\\

Hydrogen depassivation of CH-doped TMDs can generate single defects in a 2D material with atomic precision. This is a sought-after capability for defect-based quantum systems not yet demonstrated for a 2D material.
In analogy to the single transistor technologies based on hydrogen resist lithography in silicon~\cite{schofield2003atomically,he2019two}, we also envision that the dangling bonds of C$_\text{X}$ could be used as a predefined reactive docking site for other atoms or molecules. This approach will enable  embedding functional atoms in a 2D manifold in a spatially controlled manner.

\section*{Electronic and Magnetic Properties of the Carbon Radical Ion.}
\begin{figure*}[ht]
\begin{center}
\includegraphics[width=\textwidth]{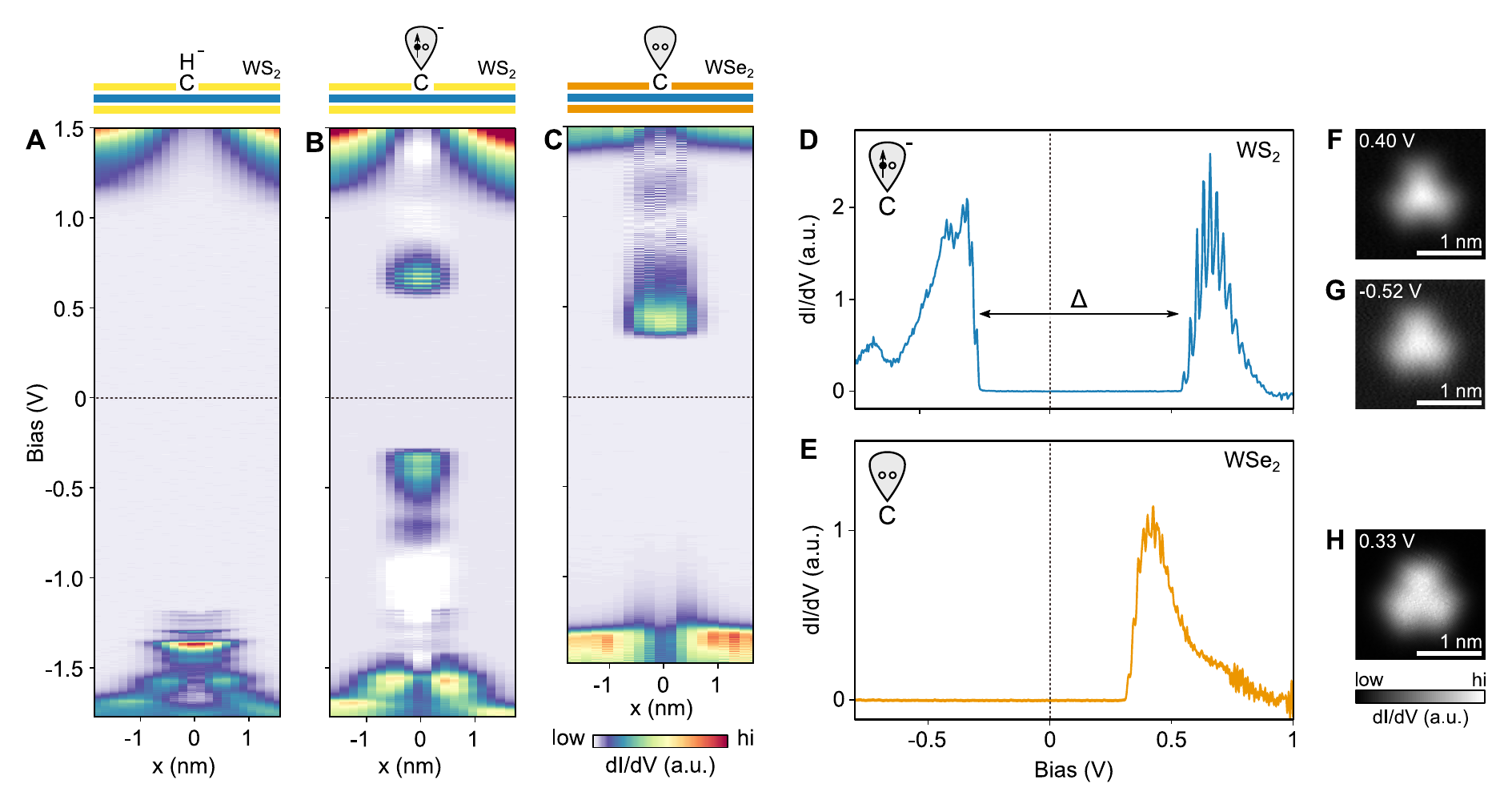}
\caption{\label{fig:dIdVacross}
\textbf{Tunneling spectroscopy of a carbon impurity in \ce{WS2} and \ce{WSe2}.}
\textbf{(A,B)} Constant height dI/dV measurement across CH$^-_\text{S}$ in monolayer \ce{WS2} before (A) and after (B) H dissociation. Both CH$^-_\text{S}$ and C$^{\bullet -}_\text{S}$ are negatively charged. The half-occupied dangling bond state of the carbon radial ion appears as two resonances in the band gap at positive and negative bias.
\textbf{(C)} Constant height dI/dV measurement across C$^0_\text{Se}$ in monolayer \ce{WSe2}. In contrast to \ce{WS2}, the carbon dangling bond in \ce{WSe2} is unoccupied and the defect is neutral. 
\textbf{(D,E)} dI/dV spectrum recorded at the center of the C$^{\bullet -}_\text{S}$ and C$^0_\text{Se}$ defect, respectively. The energy difference $\Delta = 705\,$meV between electron de- and attachment in D results from a combination of Coulomb repulsion and spin-splitting.
\textbf{(F,G)} Constant-height dI/dV maps of the C$^{\bullet -}_\text{S}$ resonance at positive and negative bias. Their similar orbital shape indicates that they originate from the same half-occupied defect state.
\textbf{(H)} Constant-height dI/dV map of the single C$_\text{Se}$ resonance.
}
\end{center}
\end{figure*}

The drastic change in STM contrast after dehydrogenation suggests a significant reconfiguration in the defect electronic structure. Scanning tunneling spectra across the CH$^-_\text{S}$ defect in monolayer \ce{WS2} are shown in \figref{fig:dIdVacross}A. The negative charge localized at the defect gives rise to a strong upwards band bending, explaining why it appears as a dark extended depression in STM images at positive bias voltage. At negative sample bias, multiple defect states are observed, which we attribute to hydrogenic bound and resonant states of the screened Coulomb potential, as we reported recently~\cite{aghajanian2020resonant}.\\

After hydrogen desorption, two prominent defect states emerge deep in the \ce{WS2} band gap, one at positive ($\sim$0.6\,V) and the other at negative ($\sim$-0.3\,V) sample bias, as seen in \figref{fig:dIdVacross}B,D. 
These highly-localized defect states are well decoupled from the dispersive bulk \ce{WS2} band structure. Each state exhibits an oscillatory fine-structure that is a signature of the vibronic coupling to the TMD lattice, which will be discussed later in detail. 
Spatial imaging of these defect states (\figref{fig:dIdVacross}F,G) reveals that they have three-fold symmetry and nearly identical orbital shapes, strongly suggesting that the two resonances originate from a single open-shell defect state~\cite{cochrane2015pronounced}. 
The gap between the defect states is slightly larger on bilayer ($\Delta = 770$\,meV) than on monolayer ($\Delta = 705$\,meV) \ce{WS2}~\footnote{These numbers were corrected for the $\sim 11\%$ voltage drop~\cite{schuler2019large} across the TMD -- Gr/SiC interface in our double-barrier tunneling junction geometry.}.
The same upwards band bending before and after dehydrogenation indicates that the defect is still negatively charged, consistent with the persisting dark halo in STM images of C$_\text{S}$.\\

In \ce{WSe2} on the other hand, the carbon defect becomes charge neutral after dehydrogenation (CH$^-_\text{Se}$ $\rightarrow$ C$^0_\text{Se}$). The charge neutrality can be deduced from the absence of band bending thus leading to a disappearance of the dark halo around the defect when imaged at positive sample biases upon dehyrogenation (see Fig.~S7), and the disappearance of associated hydrogenic states as seen in \figref{fig:dIdVacross}C. Moreover, the carbon impurity exhibits only a single, fully unoccupied defect state in the \ce{WSe2} band gap and no state at negative bias is observed (\figref{fig:dIdVacross}E). 
This neutral charge is a consequence of the different band alignments of \ce{WS2} and \ce{WSe2} with the Gr/SiC substrate. In the \ce{WSe2}/Gr/SiC heterostructure, the Fermi level lies roughly in the center of the \ce{WSe2} band gap, whereas it is \unit{443}{meV} higher for \ce{WS2}~\cite{subramanian2020tuning}. Accordingly, the underlying substrate does not donate an electron to the dangling bond-like defect state of C$_\text{Se}$ in \ce{WSe2} and thus remains empty. Alternatively, the charge state of a C\textsubscript{X} could be controlled by changing the graphene Fermi level electrostatically.~\cite{brar2011gate,brotons2019coulomb} \\

\begin{figure*}[ht]
\includegraphics[width=\textwidth]{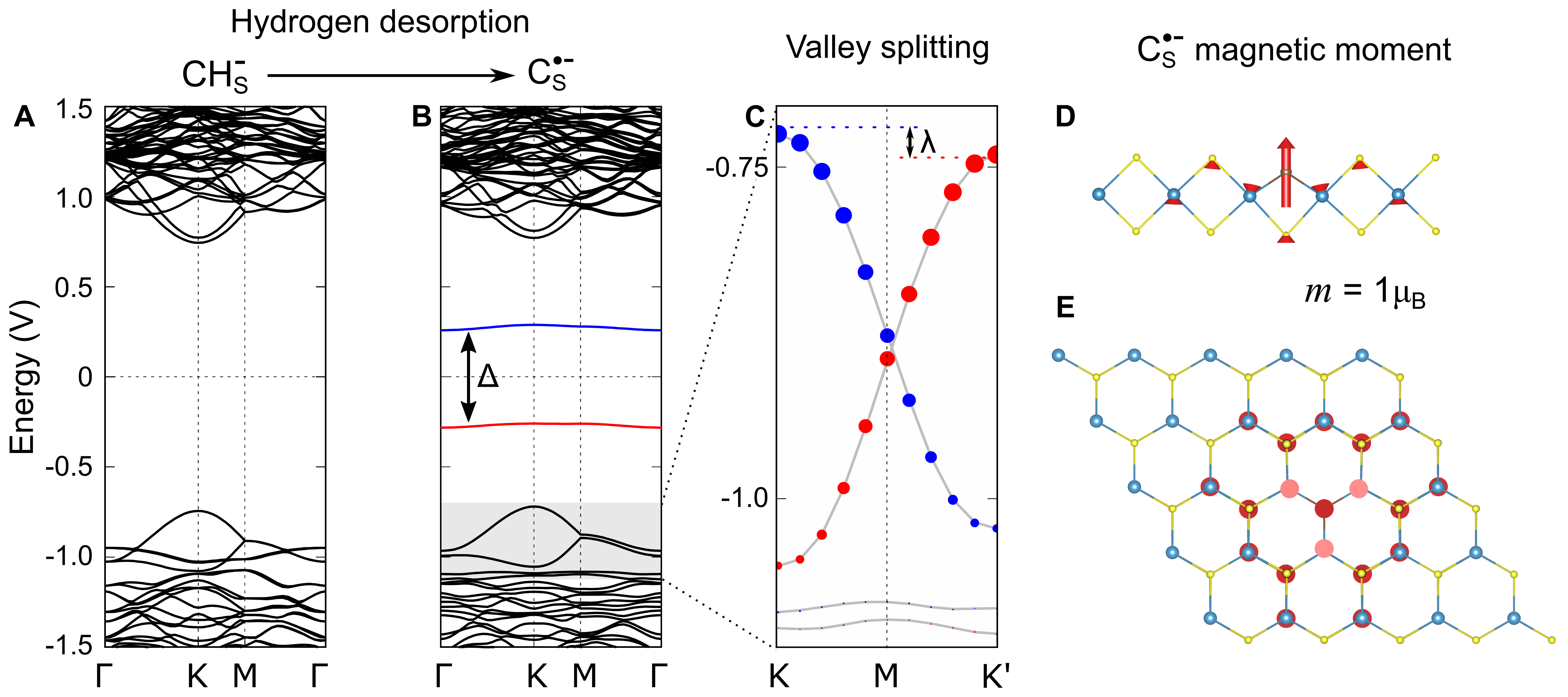}
\caption{\label{fig:theory}
\textbf{Calculated electronic and magnetic structure of a CRI.}
\textbf{(A)} Calculated band structure of the negatively charged CH$_\text{S}$ impurity using DFT-PBE in a $6\times6$ supercell with SOC. Note that in free-standing \ce{WS2} the defect would be charge neutral and the valence band half-filled. Zero energy is set to the middle of the energy gap for comparison.
\textbf{(B)} Calculated band structure of C$^{\bullet -}_\text{S}$. The corresponding density of states can be found in the Fig.~S8. C$_\text{S}$ on free-standing \ce{WS2} features only a single unoccupied defect state in the center of the \ce{WS2} band gap. Upon charging from the substrate, the now half-filled defect state is stabilized by spin-splitting ($\Delta=593$\,meV, an underestimate due to the use of DFT-PBE). The gray box marks the energy range displayed in C.
\textbf{(C)} The magnetic moment of the unpaired defect spin lifts the energy degeneracy of the K and K' valley ($\lambda=16$\,meV at a defect density of 4.7$\times10^{12}$\,cm$^2$). Red and blue colors in B and C indicate opposite spin polarization. Blue (red) dots in C represent contribution of W $l=2,m_j=-5/2(+5/2)$ state.
\textbf{(D,E)} Side and top view of the defect-localized magnetic moment of $m = 1\,\mu_\text{B}$ in the out-of-plane spin configuration.
}
\end{figure*}

Prior DFT calculations of a neutral carbon impurity in \ce{WS2} featured a single unoccupied defect state in the energy gap~\cite{zhang2019carbon}. 
Addition of one electron to form a negatively-charged carbon impurity in \ce{WS2} results in a spin polarization of the in-gap state, as shown in \figref{fig:theory}B. The total energy gain of the spin-polarized ground state as compared to the less favorable non-magnetic configuration is 147\,meV per single carbon atom (see SM for details). Hence, the negatively charged carbon impurity is characterized by a spin-split two level system of which the lower level is occupied by one electron.
Spatial maps of these defect states computed from DFT are almost identical to each other, in agreement with experimental observations (Fig.~S8). Orbital projected density of states from our DFT calculations reveal that the defect states are strongly hybridized and possess C $2p$, W $5d$, and S $3p$ orbitals character. The charged carbon impurity is computed to possess a magnetic moment of \unit{1}{$\mu_\text{B}$} from our DFT calculations, with a spin distribution closely localized at the carbon atom as shown in \figref{fig:theory}D,E. Based on these calculations, we conclude that the defect can be described as a carbon radical ion C$^{\bullet -}_\text{S}$, or CRI; and we assign the two experimentally observed in-gap states as the occupied and unoccupied spin-split defect states associated primarily with the carbon anion.\\

The defect magnetic moment can induce a spin-dependent shift of the spin-polarized \ce{WS2} valence band electrons as shown in \figref{fig:theory}C, lifting the energy degeneracy between the K and K' valleys due to time-reversal symmetry breaking. This energy shift $\lambda$ is calculated to be \unit{16}{meV} for an ordered array of defects with a density of $4.7\times10^{13}$\,cm$^{-2}$ with the PBE functional~\cite{Perdew1996Generalized} (See SM for density dependency). 
Despite the strong directional bonding of the C atom to the three neighboring W atoms below, our DFT calculations predict a weak magnetocrystalline anisotropy energy (MAE) of 0.3\,meV. This is in contrast to the giant MAE recently predicted for a Mo$_\text{S}$ antisite defect in \ce{MoS2}~\cite{khan2018room}. The vanishing MAE of C$^{\bullet -}_\text{S}$ might be expected given the small spin-orbit coupling of carbon, similar to other light element color centers.\\

The unpaired electron of the negatively charged CRI constitutes a prototypical spin-1/2 two-level system. In organic chemistry, CRIs have been studied for decades in the context of reaction intermediates. Unpaired spins of organic compounds can be detected by electron spin resonance (ESR)~\cite{russell1968electron}. Owing to their high reactivity, free radicals are usually very short-lived. In our experiments, the UHV conditions stabilizes the CRIs, but alternatively an inert capping layer such as hBN could be employed to protect the carbon dangling bond. It is also worth noting that isotopically-pure CRIs can be easily prepared by using commercially available $^{12}$C or $^{13}$C clean variants of methane in the plasma-treatment. The $^{13}$C nuclear spin offers the possibility of a local qubit memory \textit{via} the hyperfine interaction. Moreover, the low abundance of non-zero nuclear spin isotopes in certain TMDs and the intrinsically reduced spin densities in low-dimensional materials makes the TMD matrix a great host for defect spins~\cite{ye2019spin}. While spin-bath fluctuations can be expected to act favorably in 2D TMDs, electron-phonon coupling could pose another significant source of spin decoherence, which will be discussed next.\\

\section*{Vibronic Coupling of the Carbon Radical Ion.}
Each C$_\text{S}$ defect resonance above and below the Fermi energy features characteristic, equidistant peaks, a consequence of a strong electron-phonon interaction which can be probed by the transient attachment of an electron (at positive bias voltage) or a hole (at negative bias voltage) associated with the tunneling process. 
Understanding the vibronic coupling of solid-state atomic qubits is critical as it can limit the attainable coherence times~\cite{englund2010deterministic}. However, phonon sidebands can be effectively suppressed by the frequency-selective emission enhancement of a resonant cavity~\cite{englund2010deterministic,grange2017reducing}.
Moreover, low-loss local vibrations or surface acoustic phonons are analogous to nanomechanical resonators that can be used to store or coherently transmit quantum information between remote qubits~\cite{albrecht2013coupling,bienfait2019phonon,whiteley2019spin}.
Electron-phonon coupling is particularly relevant for very localized states, which in general lead to larger lattice relaxations. This applies to any deep center in wide-bandgap semiconductors. Electron-phonon interactions of defects has been extensively studied for rare-earth ions in crystals~\cite{blasse1992vibronic}, sulfur vacancies in TMDs~\cite{gupta2018franck,schuler2019large,chakraborty2020dynamic}, and a single silicon impurity in graphene~\cite{Hage2020single}.\\

In scanning tunneling spectroscopy (STS) measurements, inelastic scattering between localized charged excitations and vibrational modes is well known to lead to characteristic sideband structures. Such phenomena have been observed for molecules on surfaces~\cite{stipe1998single,franke2012effects,krane2018high}, color centers in dielectrics~\cite{repp2005scanning}, and semiconductor quantum well tunneling devices~\cite{Zou1992Inelastic}, where particular vibrational modes have been found to couple to localized electronic states. 
As mentioned in the previous section, when a CRI is introduced, we are able to clearly resolve this sideband structure in our dI/dV measurements (\figref{fig:dIdVacross}D). Interestingly this sideband structure differs substantially depending on the spin-state of the CRI, with the electron attachment (positive bias) exhibiting a clean Franck-Condon like vibronic profile while the hole attachment (negative bias) exhibits a more complex fine structure, possibly involving multiple phonons. That the vibronic structure of the spin-split defect state is so sensitive to the spin state is unintuitive given that these states derive from the same non-spin-split parent state and have the same orbital structure. 
We note that the interaction between the spin-split localized defect state and bath of harmonic phonons can be described by an effective independent spin-boson Hamiltonian~\cite{mahan2000}, as detailed in the SM. From the exact solution of this model Hamiltonian we can derive the electron spectral function $A_{\sigma}(\omega)$, that is

\begin{equation}\label{eq:spect}
A_{\sigma}(\omega) = 2\pi \sum_{\{l_1l_2...l_n\}}^{\infty} \left[ \left( \prod_{\nu=1}^{n}e^{-S_{\nu\sigma}} \frac{S^{l_\nu}_{\nu\sigma}}{l_{\nu}!}\right) \delta(\hbar\omega - \Bar{\epsilon}_{\sigma} - \sum_{\mu=1}^n\hbar\omega_{\mu}l_{\mu})\right]
\end{equation}

\noindent where $\omega_\nu$ is the frequency of the vibrational mode $\nu$, and $S_{\nu\sigma} = (g_{\nu\sigma}/\omega_\nu)^2$ is the Huang-Rhys factor, related in turn to the defect-phonon coupling strength $g_{\nu\sigma}$; and $l_{\nu}$ is an integer. Note that the Huang-Rhys factors and defect-phonon coupling strengths have a spin index, $\sigma$, while the vibrational frequencies do not, reflecting the fact that the frequencies are insensitive to the spin states while the coupling in general is not. Finally, $\Bar{\epsilon}_{\sigma}$ denotes the electronic defect state energy, including the vibrational self-energy.\\

We calculate the vibrational frequencies, $\omega_{\nu}$, using density functional perturbation theory (DFPT). Subsequently, the spin-dependent Huang-Rhys factors, $S_{\nu\sigma}$, are extracted from a spin-polarized finite-displacement DFT calculation, using eigendisplacements from DFPT. In this manner, all quantities appearing in Eq.~\eqref{eq:spect} are determined from first principles with no adjustable parameters (for details see SM).\\

\begin{figure*}[ht]
\includegraphics[width=0.5\textwidth]{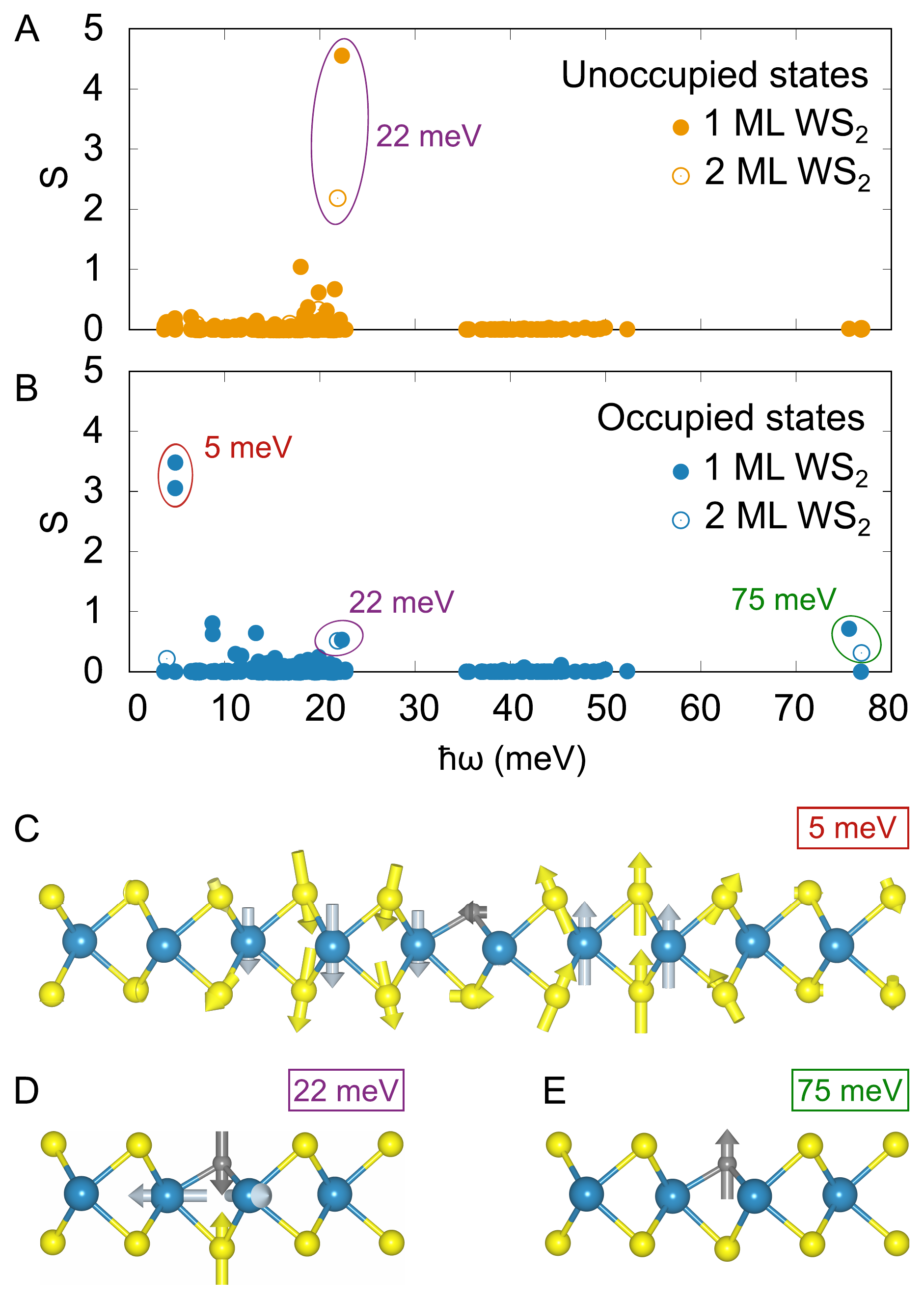}
\caption{\label{fig:VibTheory}
\textbf{Calculated electron-phonon coupling.}
\textbf{(A, B)} Calculated Huang-Rhys factors $S$ for unoccupied (A) and occupied (B) defect state for different phonons of frequency $\hbar\omega$. Filled (open) circles represent 1ML (2ML). 
The values are calculated at the $\Gamma$ point in $5\times5$, $6\times6$, and $7\times7$ supercells (shown all together). The $\Gamma$ point of these different unit cells effectively sample TMD modes with different wavevectors, which leads to convergence challenges for the low-energy resonant modes involving the defect at 5\,meV (see SM). The two local modes at 22\,meV and 75\,meV are converged for supercells larger than $5\times5$. Side projections of \textbf{(C)} the resonant mode at 5\,meV (flexural mode) and \textbf{(D, E)} the dominant local vibrational modes at 22\,meV (breathing mode) and 75\,meV (carbon out-of-plane oscillation), respectively. The experimental lattice parameter is used for all calculations. For 1ML (2ML), PBE (LDA) exchange-correlation functional is used. See SM for more details.}
\end{figure*}

\begin{figure*}[ht]
\begin{center}
\includegraphics[width=0.8\textwidth]{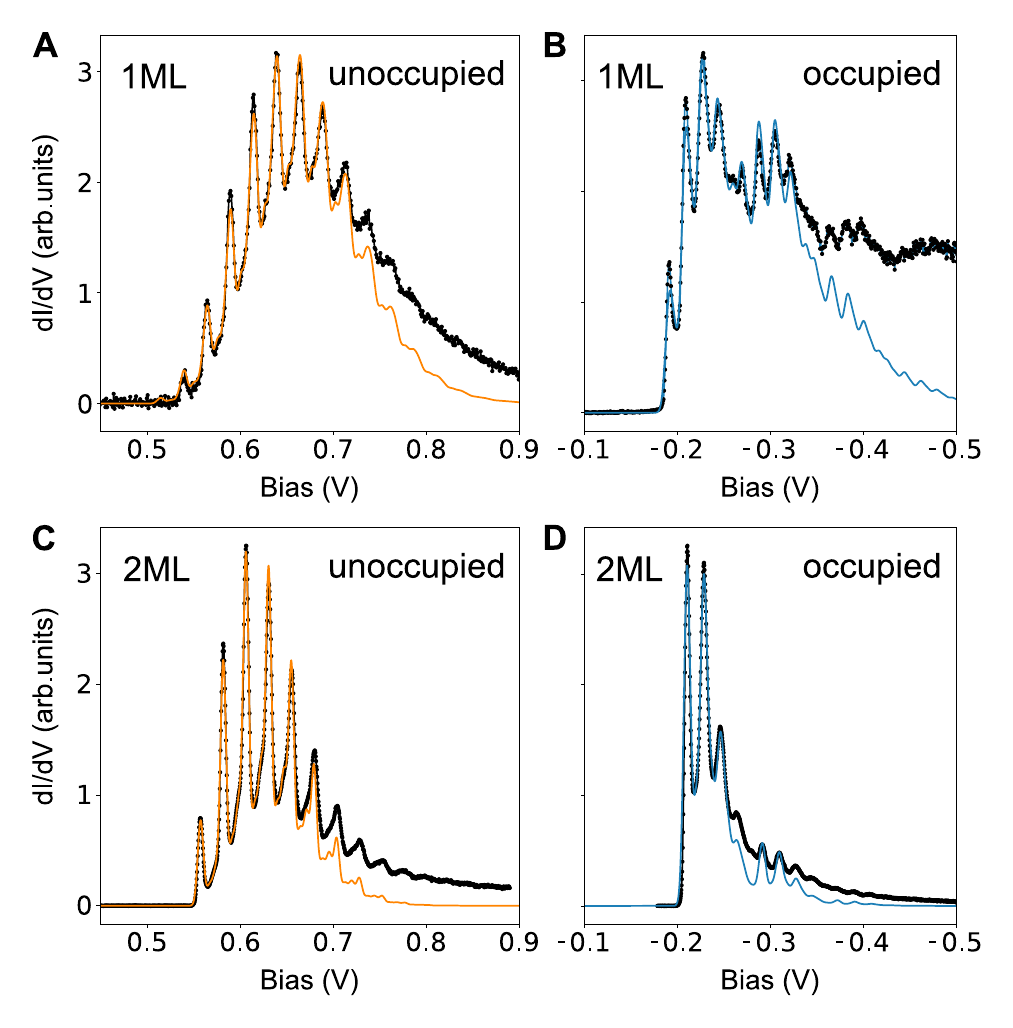}
\caption{\label{fig:vibronic}
\textbf{Vibronic excitations associated with charge state transitions of CRI.} \textbf{(A)} Electron attachment to the unoccupied defect state for C$_\text{S}$ on monolayer WS$_2$.
dI/dV measurement (black dots) and three-mode electron-phonon coupling model (orange line). 
\textbf{(B)} Hole attachment to the occupied defect state for C$_\text{S}$ on monolayer WS$_2$.
dI/dV measurement (black dots) and four-mode electron-phonon coupling model (blue line).
\textbf{(C)} Electron attachment for C$_\text{S}$ on bilayer WS$_2$. 
dI/dV measurement (black dots) and three-mode electron-phonon coupling model (orange line). 
\textbf{(D)} Hole attachment for C$_\text{S}$ on bilayer WS$_2$.
dI/dV measurement (black dots) and four-mode electron-phonon coupling model (blue line). Fit parameters are given in Tab.~\ref{tab:1}.
}
\end{center}
\end{figure*}

In our DFT calculations of a negatively-charged carbon impurity, we find two vibrational modes spatially-localized around the defect that exhibit significant coupling to the defect state, with frequencies $\hbar\omega\sim22\,$meV and 75\,meV. The 22\,meV mode is located 0.3\,meV below the top of the acoustic branch of the pristine \ce{WS2} monolayer and corresponds to an out-of-phase breathing motion involving the C-S bond and the neighboring three W atoms (\figref{fig:VibTheory}D). Interestingly, the electron-phonon coupling of this mode to the defect states is highly sensitive to spin and occupation: while we compute $S \sim 4.5$ for the unoccupied CRI state, we obtain $S \sim 0.5$ for the occupied state. This is almost an order of magnitude difference and can be attributed to the different electronic structure of the two spin-polarized defect states (see Fig.~S10). The higher energy mode around 75\,meV corresponds to a local out-of-plane C vibration (\figref{fig:VibTheory}E). The computed electron-phonon coupling of this mode is also sensitive to the specific CRI state: from our calculations, we find that it couples negligibly to the unoccupied state, $S \sim 0.01$, but moderately ($S \sim 0.7$) to the occupied state. These results are summarized in~\figref{fig:VibTheory}. As will be discussed shortly, the difference in coupling strengths are what ultimately give rise to very different sideband structure shown in \figref{fig:vibronic}A and \figref{fig:vibronic}B.\\

Repeating our calculations for a CRI defect in a \ce{WS2} bilayer identifies the same modes with significant coupling strengths as for monolayer, but with generally smaller $S$ values (see \figref{fig:VibTheory}A,B). This indicates that the defect states change less as the local vibration is excited in the bilayer system. In the bilayer, the defect states are delocalized into both layers, as shown in Fig.~S11, reducing the coupling to the lattice vibration. For the monolayer, our calculations also reveal strong coupling to low-frequency resonant flexural modes (\figref{fig:VibTheory}C) that involve the CRI defect ($\hbar\omega\sim5$\,meV), particularly for the occupied CRI state. 
We note that a hybrid acoustic-CRI defect vibrational mode is more sensitive to supercell size, increasing the uncertainty of our calculated $S$ values as detailed in the SM.

\begin{table}[h]
\centering
\begin{tabular}{lcccc}
 \toprule
  &  \multicolumn{2}{c}{\hspace{3pt} unoccupied \hspace{3pt}}  &
  \multicolumn{2}{c}{ \hspace{3pt} occupied \hspace{3pt}}\\
   & \multicolumn{1}{c}{\hspace{3pt} 1ML} \hspace{3pt} &  \multicolumn{1}{c}{\hspace{3pt} 2ML \hspace{3pt}} &  \multicolumn{1}{c}{\hspace{3pt} 1ML \hspace{3pt}}& \multicolumn{1}{c}{\hspace{3pt} 2ML \hspace{3pt}}\\
\midrule
$\hbar\omega_1$ (meV) & 10 & 10 &  6 & 10\\
$S_1$ & 0.6 & 0.2 &  0.6 & 0.3\\ 
$\hbar\omega_2$ (meV) & 19 & 16 &  \textbf{17} & \textbf{18}\\
$S_2$ &  0.7 & 0.4 & \textbf{2.2} & \textbf{0.9} \\ 
$\hbar\omega_3$ (meV) &  \textbf{25} & \textbf{24} & 26 & 23\\
$S_3$ &  \textbf{5.4} & \textbf{2.8} & 0.4 & 0.2 \\ 
$\hbar\omega_4$ (meV) &  - & - & 79 & 81 \\
$S_4$ &  - & -  & 0.7 & 0.2\\ 
$\sigma$ (meV) & 4 & 3 &  4 & 3\\
$\Bar{\epsilon}$ (meV) &  515 & 560 & -190 & -210\\
 \bottomrule
\end{tabular}
\caption{Layer and charge state dependence of the CRI vibronic coupling in \ce{WS2}. Fitted Huang-Rhys factor $S$, vibrational energy $\hbar\omega$, Gaussian broadening $\sigma$, and the electronic defect state energy $\Bar{\epsilon}$ of the three-mode (unoccupied state) and four-mode (occupied state) electron-phonon model shown in \figref{fig:vibronic}. The mode with the highest coupling strength is printed in bold.}
\label{tab:1}
\end{table}

While tempting to use the DFT values for $\omega_{\nu}$ and $S_{\nu\sigma}$ in conjunction with Eq.~\eqref{eq:spect} and compare directly with experimental STS data, we find that in practice $A_{\sigma} (\omega)$ is quite sensitive to small uncertainties in the Huang-Rhys factors. Instead, guided by the small number of phonon modes with significant coupling strength identified by theory, we fit the measured STS spectra using Eq.~\eqref{eq:spect} and subsequently compare the fit parameters to those obtained from DFT. For the unoccupied CRI state (electron attachment), three modes were sufficient for a good fit to the data; while for the occupied CRI state (hole attachment), four modes were used to largely reproduce all vibronic peaks. The fits to the tunneling spectra are shown in \figref{fig:vibronic} while the refined values for $\omega_{\nu}$ and $S_{\nu\sigma}$ used in these fits are reported in Tab.~\ref{tab:1}. All vibronic resonances exhibit a homogeneous Gaussian line broadening ($\sigma = 3\textnormal{ - }4\,$meV). Hence this broadening is likely not temperature or lifetime limited, but may be induced by coupling of the local vibrations to lattice acoustic modes.~\cite{liu2003intensity}.\\

We pause to note that the values reported in Tab.~\ref{tab:1} are not unique in that different set of $\omega_{\nu}$ and $S_{\nu\sigma}$ may be able to reproduce the STS data equally well. Thus care must be taken in interpreting the results. Nevertheless, there are some robust features independent of how the fit is performed which we can confidently compare with the DFT frequencies and couplings. Notably, we find that the pristine Frank-Condon-like sideband structure observed for the unoccupied defect state (\figref{fig:vibronic}A) derives primarily from strong defect coupling ($S \sim 5$) to a single low frequency ($\hbar\omega$ $\sim$ 20\,meV) vibrational mode. Conversely, the more complex sideband structure seen in the occupied defect-state STS spectra (\figref{fig:VibTheory}B) primarily originates from two modes, a low frequency mode ($\hbar\omega \sim $20\,meV) with moderate coupling ($S \sim 2$) and a high frequency mode ($\hbar\omega \sim $80\,meV) with a weaker coupling ($S \sim 1$), the later manifesting a beating pattern in the STS data. Remarkably, our DFT results are consistent with these results.\\

In short, we identify two vibrational modes, involving the CRI defect, that couple strongly to its defect states. While all mode energies are similar for both states and layer independent, their coupling strength is greatest for monolayer \ce{WS2} and is sensitive to the CRI defect state, indicating strong spin-phonon coupling in this system.

\section*{Summary}
In summary, we demonstrate the selective and atomically-precise generation of carbon radical ions (CRIs, C$^{\bullet -}_\text{S}$) in a TMD host crystal. In its anionic state the CRI forms a prototypical spin-1/2 system in the band gap with a magnetic moment of $1\,\mu_\text{B}$.  
Synthetically introduced CH impurities are depassivated by H desorption using a STM tip. The resulting dangling bond introduces a deep defect state in the TMD band gap that can be populated by electrons from the graphene substrate. For \ce{WS2} on Gr/SiC, the Fermi level alignment leads to a negative charge state of the C$_\text{S}$ impurity, resulting in an open-shell, spin-polarized in-gap state. 
We also demonstrate that the atomic defect couples predominantly to two vibronic modes.
While the vibrational frequencies are largely defect state and layer independent, we find that the electron-phonon coupling strength is stronger for monolayer \ce{WS2} as compared to bilayer \ce{WS2}.
The different coupling strengths to the spin-polarized defect states is a manifestation of the spin-dependent vibronic coupling in CRIs.
CRI defects in TMDs are a promising implementation of a solid-state atomic spin-qubit combining multiple favorable properties: on-demand generation with atomic precision, charge state control, tunability by external fields, surface accessibility, on-chip integration with other van der Waals materials, amenability for electrical pumping, nuclear spin memory, incorporation in a low spin-noise host, and a well-understood vibronic coupling to a low-loss local mode.

\clearpage

\section*{References}

\bibliographystyle{Science}
\bibliography{bibfinal.bib}

\section*{Acknowledgements}
We thank Andreas Schmid, Frank Ogletree and Liang Z. Tan for helpful discussions.
This work was supported as part of the Center for Novel Pathways to Quantum Coherence in Materials, an Energy Frontier Research Center funded by the US Department of Energy, Office of Science, Basic Energy Sciences.
Scanning probe measurements were performed at the Molecular Foundry supported by the Office of Science, Office of Basic Energy Sciences, of the U.S. Department of Energy under Contract No. DE-AC02-05CH11231. K.C. was supported by the University of California - National Lab Collaborative Research and Training (UC-NL CRT) program. A.W.-B. was supported by the U.S. Department of Energy Early Career Award. B.S. appreciates support from the Swiss National Science Foundation under project number P2SKP2\_171770. A.K. and J.R. acknowledge funding from Intel through the Semiconductor Research Corporation (SRC) Task 2746.001, the Penn State 2D Crystal Consortium (2DCC)-Materials Innovation Platform (2DCCMIP) under NSF cooperative agreement DMR1539916, and NSF CAREER Award 1453924. F.Z. and M.T. were supported by the Basic Office of Science of the Department of Energy under award DE-SC0018025. The authors thank Zhuohang Yu for technical support.

\section{Supplementary Materials}

\noindent Materials and Methods\\
Supplementary Text\\
Figs. S1 to S16\\

\section*{Figures}

\end{document}